\documentclass[vecphys,amsmath,amssymb]{svmult}

\usepackage{multicol}        

\usepackage{makeidx}         
\usepackage{epsfig, psfrag, graphicx}        
\usepackage{multicol}        
\usepackage[bottom]{footmisc}

 \usepackage[off,runs=2]{auto-pst-pdf}

\newcommand{\be}{\begin{equation}}
\newcommand{\ee}{\end{equation}}
\newcommand{\ben}{\begin{eqnarray}}
\newcommand{\een}{\end{eqnarray}}
\newcommand{\bea}{\begin{eqnarray}}
\newcommand{\eea}{\end{eqnarray}}
\newcommand{\bdm}{\begin{displaymath}}
\newcommand{\edm}{\end{displaymath}}
\newcommand{\ba}{\begin{align}}
\newcommand{\ea}{\end{align}}
\newcommand{\lb}{\label}

\renewcommand{\d}{\mathrm{d}}
\renewcommand{\E}{\mathrm{e}}
\renewcommand{\I}{\mathrm{i}}

\makeindex             

\begin{document}

\title*{Quantum Cosmology}

\author{Claus Kiefer\and
Barbara Sandh\"ofer}

\institute{University of Cologne, Faculty of Mathematics and Natural
  Sciences, Institute for Theoretical Physics, Cologne, Germany.} 

%
%
\maketitle



\begin{abstract}
We give an introduction into quantum cosmology with emphasis on its
conceptual parts. After a general motivation we review the formalism
of canonical quantum gravity on which discussions of quantum cosmology
are usually based. We then present the minisuperspace Wheeler--DeWitt
equation and elaborate on the problem of time, the imposition of
boundary conditions, the semiclassical approximation, the origin of
irreversibility, and singularity avoidance. Restriction is made to the framework
of quantum geometrodynamics. 
\end{abstract}

\vskip 5mm

\hspace{2.5cm}\parbox{9.2cm}{{\it For where no figure nor order is,
    there does nothing come, or go; and where this is not, there
    plainly are no days, nor any vicissitude of spaces of time.}\\
{\small Augustinus, Confessions, Liber~XII, Caput~9, transl. by E.~B.~Pusey}}


\section{Why quantum cosmology?}

Quantum cosmology is the application of quantum theory to the
Universe as a whole. 
At first glance, this may be a purely academic enterprise, since
quantum theory is usually considered to be of relevance only in the
microscopic regime. And what is more remote from this regime than
the whole Universe? This argument is, however, misleading. In fact,
quantum theory itself demands that the Universe {\em must}
be described in quantum terms. The reason is that every quantum system
except the most microscopic ones are unavoidably and irreversibly
coupled to their natural environment, that is, to a large number of
degrees of freedom that couple to the system; an example would be a small
dust grain in interaction with air molecules or photons. 
There exists then only one quantum state which describes an {\em
  entanglement} between system
and environment. The
environment is itself coupled to its environment, and so on, leading
ultimately to the whole Universe as the only closed quantum system in
the strict sense. It is this entanglement 
that leads to the classical {\em appearance} of
macroscopic bodies, a process known as decoherence. Decoherence is
well understood theoretically and has been successfully tested in a
variety of experiments \cite{deco,schlosshauer}. The Universe as a whole is thus at
the same time of quantum nature and of classical appearance
in most of its stages. There
exist, of course, also situations where the latter does not hold and
the quantum nature discloses itself; these are, in fact, the most
interesting situations, some of which we shall discuss in the course
of this article.    

Conceptually, quantum cosmology is therefore not necessarily
associated with quantum gravity. Since gravity is, however, the
dominant interaction at large scales, any realistic framework of
quantum cosmology must be based on a theory of quantum
gravity. Although there is not yet an agreement on which is the
correct theory, there exist various approaches such as quantum general
relativity and string theory \cite{oup}. The purpose of the present
article is to give a general introduction and some concrete models
which are based on quantum geometrodynamics; we make only minor
remarks on other approches such as loop quantum cosmology \cite{bojowald-book}.
 Our emphasis is on the conceptual side; for
more details we refer to \cite{oup} and the reviews
\cite{ck-isrn,coule,wiltshire,halliwell}.  

Quantum cosmology started in 1967 with Bryce DeWitt's pioneering paper
\cite{DeWitt}. He applied the 
quantization procedure to a closed Friedmann universe filled with
matter.
The latter is described phenomenologically, that is, not by a
fundamental field. 
 This is the first minisuperspace model of quantum
 cosmology. `Minisuperspace' is the generic name for a cosmological
 model with only a finite
number of degrees of freedom (such as the
 scale factor and an inflaton field). It originates from the fact that
 the full infinite-dimensional configuration space of general
 relativity is called `superspace' and the prefix `mini' is added for
 drastically truncated versions. 

DeWitt already addressed some of the important issues in quantum cosmology,
notably the singularity problem: can the singularity present in the
classical theory be avoided in quantum cosmology? 
He suggested the boundary condition that
the `wave function of the universe' $\Psi$ vanishes in the region where
the classical singularity would appear, that is, at vanishing scale factor
$a$. In fact, DeWitt strongly advocates the concept of a wave function of
the universe using the Everett interpretation and emphasizes the need
to give up the `Copenhagen' 
interpretation of quantum theory because no classical realm is
a priori present in quantum cosmology.\footnote{To quote from \cite{DeWitt}:
``Everett's view of the world is a very natural one to adopt
in the quantum theory of gravity, where one is accustomed
to speak without embarassment of the `wave function of the universe.'
It is possible that Everett's view is not only natural but essential.''} 
This coincides with modern ideas
in non-gravitational quantum theory where classical properties are
interpreted as emergent phenomena \cite{deco,schlosshauer}. Incidentally,
the linear structure of quantum theory remains untouched in almost all
papers on quantum cosmology, so an Everett-type of interpretation
is often assumed, at least implicitly. DeWitt also addressed the
important issue of the semiclassical approximation in quantum cosmology,
on which we elaborate in Section~5 below.

Quantum cosmology was soon extended to anisotropic models, notably
the Bianchi models, which are still homogeneous, but anisotropic and thus
possess different scale factors for different spatial directions.
 Reviews of this early phase in quantum cosmology 
research include \cite{misner} and \cite{ryan}.
Minisuperspace models were usually used as
illustrative examples to study conceptual issues in quantum gravity.

The second phase of quantum cosmology started in 1983 with the seminal paper
by James Hartle and Stephen Hawking on the `no-boundary proposal'
\cite{HH}. This arose from a discussion of Euclidean path integrals in
quantum gravity, which itself had its origin in black-hole thermodynamics.
The idea was to sum in the path integral over Euclidean metrics 
that only possess one boundary (the present universe) and no other, 
`initial', boundary.\footnote{For recent discussion of the status of
  this proposal, see e.g. \cite{JLM21} and the references therein.} 
Other boundary conditions include the `tunnelling condition', in which
$\Psi$ is supposed to contain only outgoing modes at singular boundaries of
superspace \cite{vilenkin} and the `symmetric initial condition' \cite{CZ}.
The issue of boundary conditions is discussed in more detail 
in Section~4 below.

After the advent of the inflationary-universe scenario around 1980,
it was also of interest to study the role of inflation in
quantum cosmology. A particular issue was the question whether
it makes sense to ask for the `probability' of inflation and, if it does,
to select the boundary condition from which inflation occurs most likely
\cite{oup,barvin}. Other issues concern the emergence of classical properties
through decoherence, the arrow of time, and the origin of structure
formation, some of which will be discussed below. Quantum cosmology
was not only discussed for models arising from the quantization of
general relativity, but also for more general situations such as
string cosmology \cite{string}
or supersymmetry (SUSY) cosmology \cite{moniz}.

The situation is different in loop quantum cosmology \cite{bojowald-book}.
A major new feature there is the occurrence of a difference
(instead of a differential) equation for the wave function of the universe.
This can lead to important results such as singularity avoidance
and the presence of a repulsive contribution to the gravitational interaction
that may be responsible for the occurrence of inflation.

In the following, we shall first introduce the general framework of
canonical quantum gravity and shall then apply it to quantum cosmology. 

\section{The formal framework: Quantum Gravity}
\lb{QG}
There are several reasons why one should try to set up a quantum
theory of gravity. The two main motivations come from quantum field
theory and general relativity, respectively. 

 From a quantum field theoretical
point of view, a unification of all fundamental interactions is an
appealing aim. This would provide quantum field theory with a
fundamental cut-off scale --- not to mention the aesthetical
aspect. Quantum gravity is then only one aspect of the ambitious
venture to unify all interactions. The most prominent representative
of such a theory so far is string theory (or M-theory). Such a unified
theory may automatically cure the divergences characteristic for local
quantum field theory. 

 From a general-relativistic perspective, a quantization of gravity is
necessary to supersede general relativity. This is due to the
remarkable fact that general relativity predicts its own
break-down. This can happen when quantities occuring in the theory 
diverge -- a type of divergence different from the quantum-field
theoretic divergences. The perhaps most important examples for such divergences are the
big-bang and black-hole singularities. More recently, other types of
singularities have received attention, mostly in connection with
models for Dark Energy; see, for example \cite{BKM19} and the
references therein. 

In our context, quantum gravity stands for an effort to provide a
quantum theory of the gravitational field.
This can be done in more or less radical ways,
depending on the amount of structure which one decides to keep
classical and at the level at which one starts to quantize the theory's
elements. The most conservative approaches are content with a
quantization of Einstein's theory of general relativity. 
Such approaches are usually subdivided into {\em covariant} and {\em
  canonical} approaches: whereas the covariant approaches employ the
covariance of spacetime at important parts of the formalism, the
canonical approaches start with a split of spacetime into space and
time and seek, in analogy to quantum mechanics, a Hamiltonian
formalism in which the four-metric is interpreted as an evolution of a
three-dimensional metric in time. Examples of covariant approaches are 
the path-integral approach and perturbation theory (derivation of
Feynman diagrams); examples of canonical approaches are quantum
geometrodynamics (the framework of this paper) and loop quantum
gravity. 

The major alternative to the direct quantization of Einstein's theory
is string theory. The ambition of this theory is to provide a unified
quantum theory of all interactions; quantum gravity is then only a
particular, ``emergent'' aspect in a limit where gravity can be distinguished as a
separate interaction. 

\subsection{Canonical Quantum Gravity}
\subsubsection{$(3+1)$-Decomposition of General Relativity}
At the basis of any canonical approach lies the Hamiltonian
formulation of general relativity, cf. \cite{oup,GK}.
To obtain such a formulation, 
one has to choose a foliation of spacetime; this is also called
$(3+1)$-decomposition.  Then
one can define a Hamiltonian on each spatial hypersurface of this
decomposition. General covariance is thus not explicit but is recovered
through the possibility to choose an arbitrary decomposition of
four-dimensional spacetime into spatial 
hypersurfaces.\footnote{As an expression is independent 
of a choice of vector basis if no
  reference to that basis is made in the expression, independence of
  the choice of foliation is maintained if the choice of foliation
  does not occur in the final equations of the theory.} 
A scheme which provides us with just such a set-up is the 
ADM-formulation of general relativity \cite{ADM,oup}. 

The peculiar feature of  general relativity is that the Hamiltonian
--- usually decomposed into a part perpendicular and three components
tangential to the spatial hypersurfaces --- is constrained to
vanish (We restrict ourselves to closed cosmologies;
  otherwise the Hamiltonian may contain boundary terms.). It is a linear
combination of the four local constraints
\be
\mathcal{H}_\perp({\mathbf x})=0\ ,\qquad \mathcal{H}_a({\mathbf x})=0\ ,
\ee
where $a=1,2,3$, called the Hamiltonian and momentum (diffeomorphism)
constraints, respectively. 
These constraints cover the entire dynamics of
Einstein's theory and are equivalent to the Einstein equations.
The occurrence of constraints is due to the fact that general relativity is a
diffeomorphism-invariant theory (loosely speaking, this is the
invariance under coordinate transformations).
 Figuratively speaking,
diffeomorphism-invariance implies that spacetime points themselves
cannot be endowed with any meaning in general relativity.

Canonical coordinates in this formalism are the 
three-dimensional metric, $h_{ab}$, on the spatial hypersurface
and its conjugate momentum, related to the extrinsic
curvature and denoted by
$p^{ab}$. A different choice of canonically-conjugate
  coordinates is made in loop quantum gravity,
  cf. \cite{oup,rovelli,bojowald-book}. The
  $(3+1)$-decomposition is illustrated in Figures 1 (general
  case) and 2 (cosmological example).

\begin{figure}
\begin{minipage}[t]{0.45\linewidth}
\scalebox{1.0}{\hspace{-10mm}\includegraphics[angle=0]{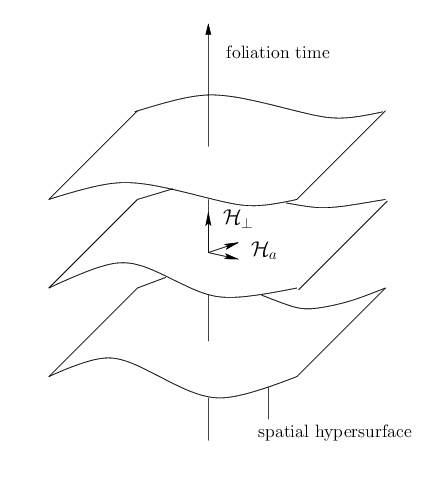}}
\caption{\lb{ADM} The above plot shows the $(3+1)$-decomposition of a
  four-dimensional spacetime. Spatial hypersurfaces are stacked
  together along a foliation parameter. The components of the
  Hamiltonian tangential and perpendicular to the hypersurfaces are
  shown (but note that there are actually three components tangential
  to the hypersurfaces).}
\end{minipage}
\hfill
\begin{minipage}[t]{0.45\linewidth}
\scalebox{1.0}{\hspace{-10mm}\includegraphics[angle=0]{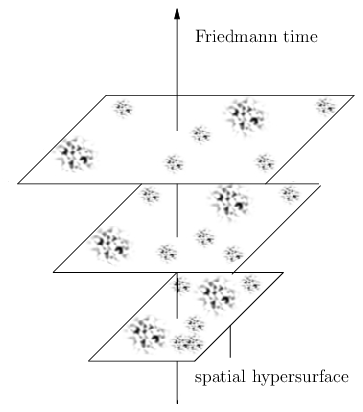}}
\caption{\lb{Universe} Here, the decomposition is illustrated with
  the help of a cosmological example. The foliation is chosen such
  that spatial hypersurfaces are homogeneous and isotropic. On each
  such hypersurface we see galaxy clusters.}
\end{minipage}
\end{figure}

\subsubsection{Quantization of the constraint system}
Quantization can be carried out in analogy to ordinary quantum
theory: canonical coordinates are promoted to operators, thus turning
the constraints into operators. These are implemented through the
requirement
\be
\widehat{\mathcal{H}}_\perp\Psi[h_{ab}]=0\ ,
\qquad \widehat{\mathcal{H}}_a\Psi[h_{ab}]=0\ ,
\ee
where $\Psi[h_{ab}]$ is a wave functional depending on the
three-metric. If non-gravitational fields are present, the wave
functional depends in addition on these other fields, for example
a scalar field $\phi$.
Here and in the following, operators will be denoted by hats on the
corresponding quantities.    
The first equation is usually referred to as Wheeler--DeWitt
equation. The momentum constraints
ensure invariance of the wave functional under
infinitesimal three-dimensional diffeomorphisms.  

In the classical theory, spacetime can be represented as a foliation
of three-dimensional hypersurfaces. Since the canonical variables are
the three-dimensional metric and the embedding of the hypersurfaces
into the fourth dimension, both variables cannot be specified
simultaneously in the quantum theory (they obey an `uncertainty
relation') -- spacetime has disappeared, in the same manner as the
classical trajectories have disappeared in quantum mechanics. What remains is the
configuration space: the space of all three-geometries (called
`superspace' by John Wheeler). For more details on superspace and its
construction, see \cite{Giulini09}. 

What has been written down in a formal way here is in fact the origin of
ambiguities in the set-up of a quantum theory of gravity. Promoting
canonical coordinates to operators requires a choice of
representation. And it is exactly at this point where the canonical
path to quantum gravity branches out, yielding loop quantum gravity 
and quantum geometrodynamics as the most prominent directions. We will follow
here the geometrodynamical path being given by the choice of
Schr\"odinger representation of fundamental operators.
In geometrodynamics, then, the three-metric acts a multiplication
operator and the momentum as a derivative operator, 
\be
\widehat{h}_{ab}\Psi[h_{ab}]=h_{ab}(\vec{x})\cdot\Psi[h_{ab}]\ ,
\qquad
\widehat{p}_{cd}\Psi[h_{ab}]=\frac{\hbar}{\mathrm{i}}\frac{\delta\Psi}{\delta
h_{cd}(\vec{x})}[h_{ab}]\ ,
\ee 
where ${\mathbf x}$ denotes a point in the hypersurface.
With this choice of representation, the governing quantum equations
read, in the vacuum case,
\ben
\lb{fullwdw}
\left(-16\pi G\hbar^2G_{abcd}\frac{\delta^2}{\delta h_{ab}\delta
    h_{cd}}-\frac{\sqrt{h}}{16\pi G}(^{(3)}R-2\Lambda)\right)\Psi=0\ ,\\
-2D_bh_{ac}\frac{\hbar}{\mathrm{i}}\frac{\delta\Psi}{\delta h_{bc}}=0\ .
\een
In tne non-vacuum case, these are supplemented by terms coming from
the respective components of the energy--momentum tensor,
$T_{00}$ and $T_{0i}$.
These equations are of a formal nature, since the factor-ordering
problem and the regularization issue have not been addressed. (A recent
example for a regularization can be found in \cite{Feng18}.) 
Here, $G$ is the gravitational constant and $\hbar$ Planck's
constant, $\mathrm{i}$ is the imaginary unit.\footnote{The speed of
  light is set equal to one. It can be restored by the substitution
  $G\to G/c^4$.}
The first, Wheeler--DeWitt, equation contains two terms. 
The first term
with the structure of a kinetic energy term contains second
derivatives with respect to the metric on spatial hypersurfaces,
$h_{ab}$. They are connected through the so-called DeWitt metric
$G_{abcd}$. The second term, acting as a potential, enters with a factor
containing the square root of the determinant of the
three-metric, $h$. It has two contributions, one coming from the
three-dimensional Ricci scalar $^{(3)}R$ and, counteracting this, the
cosmological constant $\Lambda$. 
The diffeomorphism constraint equations contain a covariant
derivative $D_b$ of the $3$-metric. The equation is of a similar form as the
Gauss constraint in quantum electrodynamics.
Even though most conceptual features of this set of constraint
equations can be discussed at the fundamental level \cite{oup}, we will now turn
to cosmological examples to illustrate them.

\section{Wheeler--DeWitt equation}
\lb{WDW}
In homogeneous cosmologies, the diffeomorphism constraint is satisfied
trivially. What remains is the Wheeler--DeWitt equation, which is at
the heart of any canonical approach to quantum cosmology, independent
of the choice of representation.\footnote{In loop quantum cosmology it
takes the form of a difference equation.} It is the analogue of the
(functional) Schr\"odinger equation in quantum field theory.

To be more precise, consider the simplest setting, a
homogeneous, isotropic Friedmann cosmology filled with a perfect
cosmological fluid. This is an example of a `minisuperspace model'.
As we are aiming at a quantum theory, instead of an
effective description of the cosmological fluid in terms of its energy
density and pressure, we use a fundamental Lagrangian, namely that of
a scalar field. The scalar field thus serves as a surrogate for the
matter content of the universe. Our fundamental equation is then given
by (see the Appendix for a derivation)
\bea
\lb{whdw1}
\hat{\mathcal H}\Psi&=&\left(
\frac{2\pi G\hbar^2}{3}\frac{\partial^2}{\partial\alpha^2}-
\frac{\hbar^2}{2}\frac{\partial^2}{\partial\phi^2}\right.\nonumber\\ & & +
\left.a_0^6\E^{6\alpha}\left(V\left(\phi\right)+\frac{\Lambda}{8\pi G}\right)
-3a_0^4\E^{4\alpha}\frac{k}{8\pi G}\right)\Psi(\alpha,\phi)=0\ ,
\eea
with cosmological constant $\Lambda$ and curvature index $k=\pm 1,
0$.
The variable $\alpha=\ln a/a_0$, where $a$ stands for the
scale factor and $a_0$ is a reference scale, is introduced to obtain a convenient form of the
equation.  

As it stands, this is actually an understatement of the
  importance of this step. The introduction of the variable $\alpha$
  is necessary to endow the Wheeler--DeWitt equation with a quantity 
which ranges from negative to positive infinity. Indeed the r\^ole of the
  variable $\alpha$ is two-fold: it leaves us with positive values for the scale factor 
 {\it and} makes the Wheeler--DeWitt equation
 well-defined at $a\to 0$. This can be seen from the original
  form of the Wheeler--DeWitt equation containing $a$,
\bea
\lb{whdw2}
\hat{\mathcal H}\Psi&=&\left(
\frac{2\pi G\hbar^2}{3a^2}\frac{\partial}{\partial
  a}\left(a\frac{\partial}{\partial a}\right)-
\frac{\hbar^2}{2a^3}\frac{\partial^2}{\partial\phi^2}\right.\nonumber\\ & & +
\left.a^3\left(V\left(\phi\right)+\frac{\Lambda}{8\pi G}\right)
-3a\frac{k}{8\pi G}\right)\Psi(a,\phi)=0\ .
\eea
This equation obviously contains terms that diverge as the big bang 
(or the big crunch) is approached, $a\to 0$. 
Through a transformation on $\alpha$ we obtain
(\ref{whdw1}) multiplied by an overall factor of $\E^{-3\alpha}$. But this factor
is non-zero and can be removed, leaving the well-defined equation
(\ref{whdw1}).\footnote{The attempts to implement the positivity of
  the metric into the fundamental commutation relations leads to the
  approach of `affine quantum gravity' \cite{klauder}. As far as
  quantum cosmology is concerned, this approach is very close to the
  approach presented here.}

Quantization of the Hamiltonian constraint thus provides quantum
cosmology with a central equation of equal importance as the
Schr\"odinger equation in quantum mechanics. 
But there are some obvious differences to the
 Schr\"odinger equation which cannot go without comment and which will
 be discussed in the following subsections. 

\subsection{An equation in configuration space}
Most obviously, the Wheeler--DeWitt equation is a partial differential
equation determining a wave function which is {\it not} defined in
space or time or spacetime. It is an equation defined in
configuration space, that is, it depends on the gravitational and matter
degrees of freedom of the system (in this example $\alpha$ and $\phi$)
 --- not on spacetime
points. This is in fact what one would expect from the quantization of a
diffeomorphism-invariant theory such as general relativity. In general
relativity, a spacetime point has no physical significance. It can be
assigned significance only through a physical field. That is why we
find not points but fields as the arguments of the wave function. 
\subsection{A timeless equation}
Furthermore, this equation lacks an external time parameter. If we
think of the Wheeler--DeWitt equation as the analogue of the
Schr\"odinger equation, the most striking difference is the lack of a 
first derivative with an imaginary factor. This is what, in
quantum mechanics, distinguishes space from time: time $t$ is
represented by a real number, and the positions are represented by
operators. In special relativistic quantum field theory, spacetime
plays the role of the external time, and the dynamical quantum fields
are represented by operators. In quantum gravity, spacetime has
disappeared and only the quantum fields (defined on space) remain.

Apart from the
observation that we have no coordinates and thus also no
time-coordinate dependence in the Wheeler--DeWitt equation, we find
that we have no preferred evolution parameter. This is usually called
the {\em problem of time}: 
the absence of an external time parameter
and the non-uniqueness of internal timelike variables. This fact has
caused much confusion at the
advent of canonical quantum gravity.\footnote{For a recent essay on
  time in quantum cosmology, see \cite{KP21} and the references therein.} 
There are basically two types of reaction to this observation.

\paragraph{Time before quantization --- internal time}
One possibility to cope with this situation is to try to recover the
form of the Schr\"odinger equation in some way. The basic idea is to 
rewrite the classical constraints in such a way that upon
quantization, a Schr\"odinger-type equation is obtained. At the
classical level, the rewritten constraints thus have to be of the form
\be
\lb{SolvedHamiltonian}
\mathcal{P}_A+h_A=0\ ,
\ee
that is, one needs a linear canonical momentum $\mathcal{P}_A$ for which
one can solve the constraint; $h_A$ simply stands for the remaining terms.
Upon quantization one obtains
\be
\mathrm{i}\hbar\frac{\delta\Psi}{\delta q_A}=\widehat{h}_A\Psi\ ,
\ee
which is a Schr\"odinger-type 
equation. It is,
  in general, inequivalent to the Wheeler--DeWitt equation.
Usually, $h_A$ is referred to as `physical Hamiltonian' because it
actually describes an evolution in a {\it physical} parameter, namely the
coordinate $q_A$ conjugate to $\mathcal{P}_A$. Time, in the sense of a
physical evolution parameter, is therefore defined {\it before}
quantization. As a result, time is part of an a priori background
structure.
 
Primary research in this direction tried to actually solve the
constraints of general relativity, separating the true, dynamical
from the gauge degrees of freedom. This attempt to cope with the
timelessness of quantum gravity suffers from several conceptual as
well as technical difficulties; for an extensive list,
see \cite{Isham,kuchar,oup,anderson-book}.
An example of a conceptual problem is the fact that the choice of time
variable is not unique. It is not clear what conditions a variable has
to satisfy in order to qualify as an internal time. Neither is it
clear that, could such conditions be specified, they would leave a
unique choice of time.

In the simple cosmological case, one could, for example,
choose the scale factor but equally well the scalar field $\phi$
as internal time. But one could as well choose one of the canonical
momenta, for example the one related to the scale factor, $\pi_a$, as
internal time coordinate. From these choices result non-unitary,
that is, inequivalent, quantum theories. It is unclear how predictions
resulting from different such choices are related (if they are related
at all).

Technically, the heaviest blow came from a result by Charles
 Torre who proved that a global solution to the constraints does not exist
in general relativity \cite{Torre}.
In consequence, advocates of this interpretation of the
constraint equations turned to `matter clocks' \cite{kuchar}. 
A matter clock is a type of
matter whose Hamiltonian is of such a form that the full Hamiltonian
is again of the form (\ref{SolvedHamiltonian}); that is, the matter
Hamiltonian has to be linear in momentum and to describe physical clocks
(e. g. they should run forward). Matter clocks in the form of a scalar
field are frequently used in loop quantum gravity \cite{bojowald-book}.

\paragraph{Time after quantization --- the frozen formalism} 
An opposite strategy is to identify time {\it after} quantization.  Here,
the Wheeler--DeWitt equation is taken at face value and no rewriting
at the classical level is made before quantization. The crucial
observation here goes back to DeWitt  who accentuated the (local)
hyperbolic form of the Wheeler--DeWitt 
equation \cite{DeWitt}, see also \cite{Giulini09}.\footnote{Thus, in
  structure, the Wheeler--DeWitt equation resembles the
  Klein--Gordon equation rather than the Schr\"odinger
  equation. Consequently, it suffers from the same problems as the
  former one (lack of positive-definiteness of the inner product).} The
hyperbolic form distinguishes a particular part of 
the gravitational degree of freedom (its conformal part, which in the
above model is just the scale factor $a$) from
the remaining degrees of freedom: its second-derivative term has a positive
prefactor, whereas the derivatives with respect to the remaining degrees
of freedom enter with a minus sign. In this spirit, 
it seems sensible to impose boundary
conditions for fixed gravitational degree of
freedom, $a=\mathrm{constant}$.

Problems of this point of view are all related to the lack of
time structure. There seems to be
 no way to define a positive definite inner product
in general models. Thus we have no handle on the interpretation of the
wave function. Related to this is the question whether operators,
for example the constraint operators, should be self-adjoint.
But all these problems which at the dawn of quantum
gravity drove researchers to the internal-time idea, are due to the
fact that we have no a priori notion of time. As our understanding of
quantum gravity has improved, we may be able to challenge our brain
with this further loss of structure. Inner products and self-adjoint
operators are then reserved for a semi-classical world where a notion
of time can be recovered, see Section \ref{Inhomogeneities}. 
In the following we will adopt this point of view. 

\subsection{Determinism of wave packets}
The two rather straight observations made above concerning the
structure of the Wheeler--DeWitt equation converge to producing a
peculiar notion of determinism at the level of quantum cosmology.
Despite
the absence of an {\it external} time parameter, the equation is of
hyperbolic form thus suggesting to use the $3$-volume $v$ or
$\alpha=\frac13\ln v$ as an {\it intrinsic} time parameter \cite{Zeh88}.
The term `intrinsic time parameter' denotes an evolution parameter 
of the equation, generally unrelated to any physical notion of time 
(which at the quantum level is anyway lost, as discussed above).
Exchanging the
classical differential equations in time for a timeless differential
equation hyperbolic in $\alpha$ alters the determinism of the theory. 
This, of course, changes
the way in which boundary conditions can be imposed. Wave packets are
not evolved with respect to Friedmann time but with respect to intrinsic
time. This turns our notion of determinism on the head. 

Let us illustrate this by the following example. Simplify the
universe model with the two degrees of freedom $a$ (scale factor) and
$\phi$ (scalar field)
underlying (\ref{whdw2}) by the assumption
$\Lambda=0$. Take, moreover, the scalar field to be massless and the
universe to be closed, $k=1$. This model has a classical solution
evolving from big bang to big crunch. The trajectories in
configuration space are depicted in Figure~3 where the arrow
along the trajectory signifies increasing Friedmann time.  

\begin{figure}[h]
\label{fig_berlin05_6ab}
\caption{The classical and the quantum theory of gravity exhibit
drastically different notions of determinism. The scalar field $\phi$ is
shown on the horizontal axis, while the scale factor $a$ of the
universe is shown on the vertical axis.}
\begin{center}
\includegraphics[width=5cm]{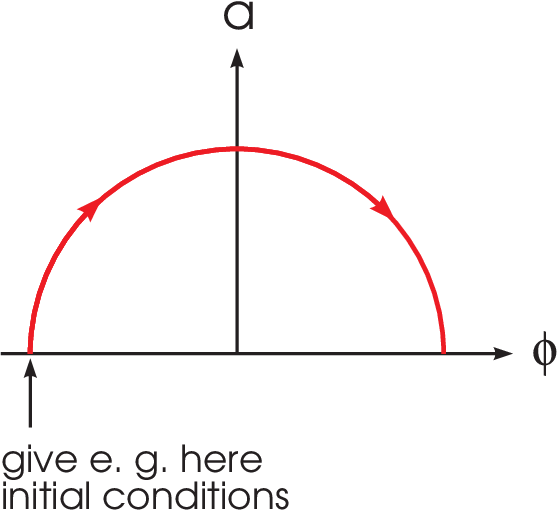}
\includegraphics[width=5cm]{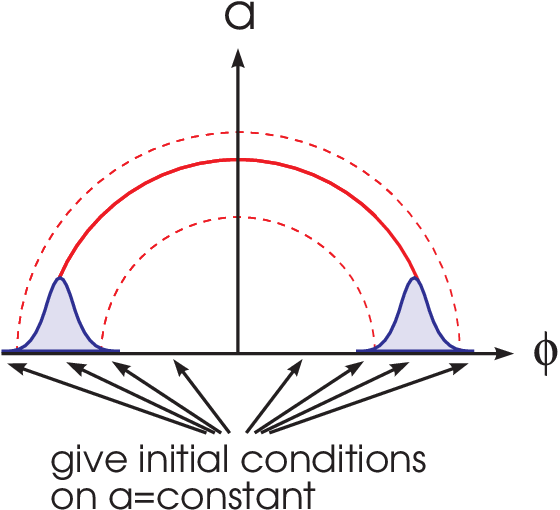}
\end{center}
\end{figure}

Classically, one imposes initial conditions at
$t=t_0$, corresponding to the left intersection of the trajectory with
the $\phi$-axis. These initial conditions determine the
evolution of $a$ and $\phi$ into the big-crunch singularity. Not so in
quantum cosmology. Here, initial conditions have to be imposed along
the whole axis
$a=$ constant. If the wave packet shall follow the classical trajectory, one
has to impose two wave packets at each intersection point of the
classical trajectory with the $a=$ constant line. Wave packets are evolved from both,
classical big-bang and big-crunch singularity, in
the direction of increasing $a$; big bang and big crunch are
intrinsically indistinguishable.\footnote{
One should remark that the hyperbolicity with respect to the
three-volume is an inherent feature of the Wheeler--DeWitt equation only 
as long as we
are working in physically conventional settings. A counter-example is
given by a non-minimally coupled scalar field. In this case, regions in
configuration space exist in cosmological models 
where the quantum equation becomes
elliptic \cite{CK}. Moreover, implementation of more exotic types of matter may
also destroy the hyperbolicity of the Wheeler--DeWitt equation. 
Including, for example, a
phantom field mimicked by a homogeneous scalar field with reversed
sign of the kinetic energy term, yields an ultrahyperbolic equation
\cite{MCB,BKM19}. An
evolution of initially imposed wave packets at $a=\mathrm{constant}$ is
not justified in these cases by the form of the Wheeler--DeWitt equation.} 
\section{Boundary Conditions}
\lb{BoundaryConditions}
As we have discussed above, implementing boundary conditions in
quantum cosmology differs from the situation in both general
relativity and ordinary quantum mechanics. In the following, we shall
briefly review two of the most widely discussed boundary conditions:
the `no-boundary proposal' and the `tunnelling proposal'. More details
can be found in \cite{oup}.

\subsection{No-boundary proposal}

Also called the `Hartle--Hawking proposal' \cite{HH}, the no-boundary
proposal is essentially of a topological nature. It is based on the
Euclidean path integral representation for the wave function,
\begin{equation}
\label{Zg}
\Psi[h_{ab}]=\int{\mathcal D}g_{\mu\nu}(x)\, 
{\rm e}^{- S[g_{\mu\nu}(x)]/\hbar}\ ,
\end{equation}
in which $S$ is the classical action of general relativity and
${\mathcal D}g_{\mu\nu}(x)$ stands for the integration measure - a sum
over all four-geometries. (In
general, one also sums over matter fields.) `Euclidean' means that the
time variable is assumed to be imaginary (`imaginary time').
 
Since the full path integral cannot be evaluated exactly, one usually
resorts to a saddle-point approximation in which only the dominating
classical solutions are taken into account to evaluate $S$.
The  proposal, then, consists of two parts. First, it is assumed
that the Euclidean form of the path integral is fundamental, and that the
Lorentzian structure of the world only emerges 
in situations where the saddle point is
complex. Second, it is assumed that one integrates over metrics with
one boundary only (the boundary corresponding to the present universe),
so that no `initial' boundary is present; this is the origin of the
term `no-boundary proposal'. The absence of an initial boundary
is implemented through appropriate regularity conditions.   
In the simplest situation, one finds the dominating geometry 
depicted in Figure~4, which is often called the `Hartle--Hawking
instanton', but which was already introduced by Vilenkin \cite{plb82}:
the dominating contribution at small radii is (half of the) Euclidean
four-sphere $S^4$, whereas for bigger radii it is (part of) de Sitter space,
which is the analytic continuation of $S^4$. Both geometries are matched
at a three-geometry with vanishing extrinsic curvature.
The Lorentzian nature of our universe would thus only be an `emergent'
phenomenon: standard time $t$ emerges only during the `transition'
from the Euclidean regime (with its imaginary time) to the Lorentzian regime.  

\begin{figure}[h]
\caption{`Hartle--Hawking instanton': the dominating contribution to the
Euclidean path integral is assumed to be half of a four-sphere attached
to a part of de Sitter space. Obviously, this is a singularity-free
four-geometry. This instanton demonstrates clearly the no-boundary
proposal in that there is no boundary at $\tau=0$.}
\begin{center}
\includegraphics[width=6cm]{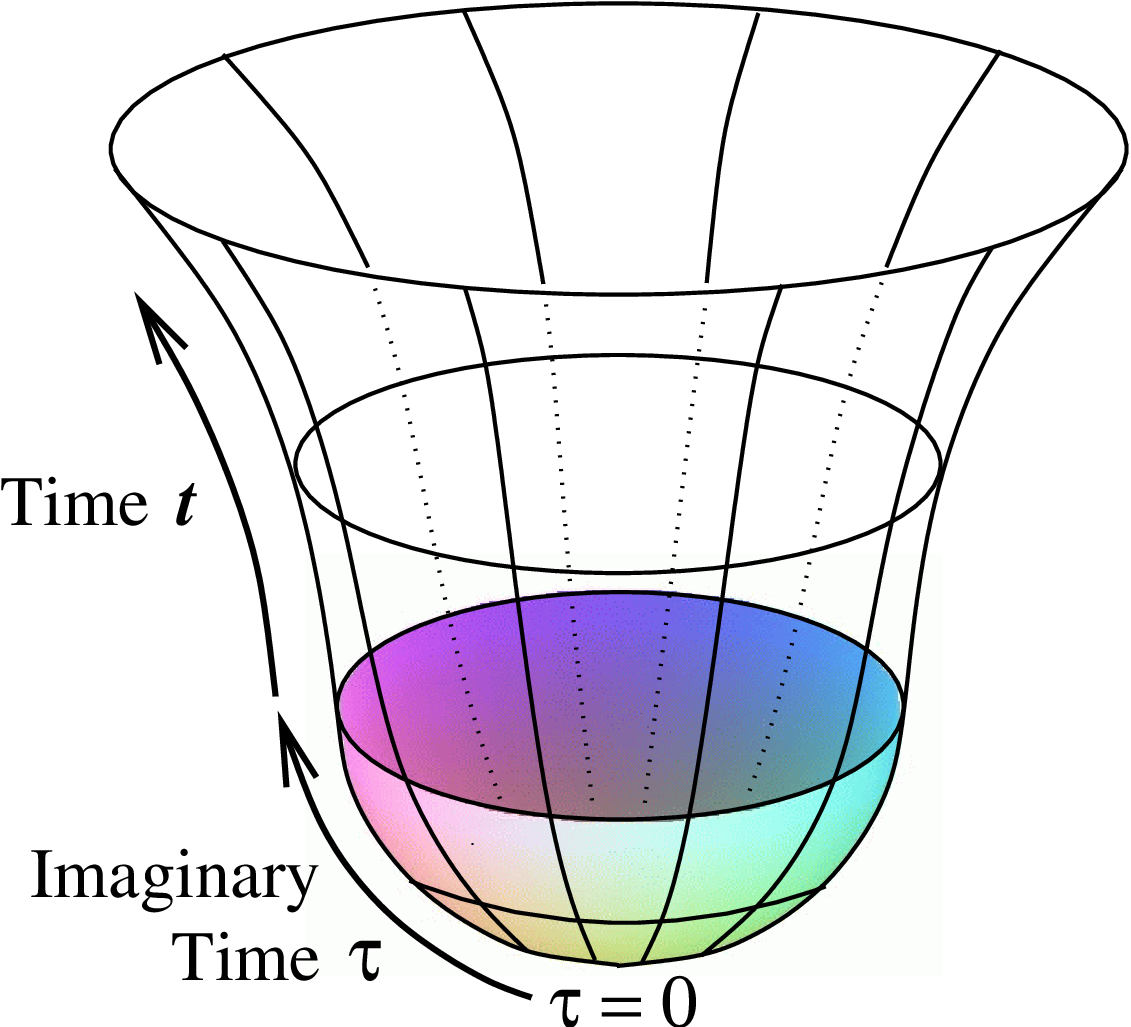}
\end{center}
\end{figure}

 From the no-boundary proposal one can find for the above model with
 the massive scalar field (and vanishing $\Lambda$) the following wave
 function in the Lorentzian regime:
\be
\label{NB}
\psi_{\rm NB}\propto \left(a^2V(\phi)-1\right)^{-1/4}
\exp\left(\frac{1}{3V(\phi)}
\right)\cos\left(\frac{(a^2V(\phi)-1)^{3/2}}{3V(\phi)}-\frac{\pi}{4}\right)\ .
\ee

In more general situations, one has to look for integration contours
in the space of complex metrics that render the integral convergent.
In concrete models, one can then find a class of wave functions which
is a subclass of the solutions to the Wheeler--DeWitt equation. In
this sense, the boundary condition picks out particular
solutions. Unfortunately, the original hope that only one definite
solution remains, cannot be fulfilled.

One questions that arises is whether one can recover from the
no-boundary wave functions with correct quasi-classical behaviour,
that is, wave functions from which one can construct wave packets
following the classical trajectories. Using an exactly soluble model,
it was shown that this is in general not the case
\cite{kiefer91}. Instead, one finds solutions that either diverge at
infinity or along the ``light cones'' defined by the DeWitt metric.
More recent investigations also point to problems of the no-boundary
proposal regarding the classical limit \cite{MT20}.
In addition, there is an ongoing discussion about questions of
stability; see, for example,
\cite{JLM21} and the references therein. The status of the no-boundary
proposal thus remains open. 

\subsection{Tunnelling proposal}

The tunnelling proposal emerged from the work by Alexander
Vilenkin, cf. \cite{plb82,vilenkin} and references therein. It
is most easily formulated in mini\-superspace. 
In analogy with, for example, the process of $\alpha$-decay in quantum
mechanics, it is proposed that the wave function consists solely of
{\em outgoing} modes. More generally, it states that it consists solely
of outgoing modes at singular boundaries of superspace (except the
boundaries corresponding to vanishing three-geometry). In the
minisuperspace example above, this is the region of infinite $a$ or
$\phi$. What does `outgoing' mean? The answer is clear in quantum mechanics,
since there one has a reference phase $\propto\exp(-\I\omega t)$. An
outgoing plane wave would then have a wave function
 $\propto\exp(\I kx)$. But since there
is no external time $t$ in quantum cosmology, one has to {\it define}
what `outgoing' actually shall mean \cite{Zeh88,conradi}. Independent of
this reservation, the tunnelling proposal picks out particular
solutions from the Wheeler--DeWitt equation. The interesting fact is
that these solutions usually differ from the solutions picked out by
the no-boundary proposal: whereas the latter yields real solutions,
the solutions from the tunnelling proposal are complex;
 the real exponential prefactor differs in the sign of
the exponent. Explicitly, one gets in the above model the following
wave function:
\be
\psi_{\rm T}\propto (a^2V(\phi)-1)^{-1/4}\exp\left(-\frac{1}{3V(\phi)}
\right)\exp\left(-\frac{\I}{3V(\phi)}(a^2V(\phi)-1)^{3/2}\right)\ .
\ee
Comparing this with (\ref{NB}), one recognizes that the tunnelling
proposal leads to a wave function different from the no-boundary condition.
Consequences of this difference arise, for example, if one asks for
the probability of an inflationary phase to occur in the early
universe: whereas the tunnelling proposal seems to favour the occurrence
of such a phase, the no-boundary proposal seems to disfavour it.
An example where the tunnelling proposal can successfully predict
inflation with a non-minimally coupled Standard Model Higgs field is
discussed in \cite{BKKS10}.

\section{Inclusion of inhomogeneities and the semiclassical picture}
\lb{Inhomogeneities}

Realistic models require the inclusion of further degrees of freedom;
after all, our Universe is not homogeneous. This is usually done by adding
a large number of multipoles describing density perturbations and
small gravitational waves \cite{HH85,oup}. One can then derive an
approximate Schr\"odinger equation for these multipoles, in which the
time parameter $t$ is defined through the minisuperspace variables
(for example, $a$ and $\phi$). The derivation is performed by a
Born--Oppenheimer type of approximation scheme.
The result is that the total state (a solution of the Wheeler--DeWitt
equation) is of the form
\begin{equation}
\label{expiS}
\Psi \approx \exp({\rm i}S_0[h_{ab}]/\hbar) \, \psi[h_{ab},\{ x_n\}]\ ,
\end{equation}
where $h_{ab}$ is here again the three-metric, $S_0$ is a function of
the three-metric only, and
$\{ x_n\}$ stands for the inhomogeneities (`multipoles').
 In short, one has that
\begin{itemize}
\item $S_0$ obeys the Hamilton--Jacobi equation for the gravitational field
and thereby defines a classical spacetime which is a solution to
Einstein's equations (this order is formally similar to the recovery
of geometrical optics from wave optics via the eikonal equation).
\item $\psi$ obeys an approximate Schr\"odinger equation,
\begin{equation}
\label{semi}
 {\rm i}\hbar \, \underbrace{\nabla \, S_0 \, \nabla
\psi}_{\equiv\frac{\displaystyle\partial\psi}{\displaystyle\partial t}} 
\approx H_{\rm m} \, \psi \ ,
\end{equation}
where $H_{\rm m}$ denotes the Hamiltonian for the multipole degrees of
freedom. The $\nabla$-operator 
on the left-hand side of (\ref{semi}) 
is a shorthand notation for derivatives with respect to the
minisuperspace variables (here: $a$ and $\phi$). 
Semiclassical time $t$ is thus defined in this limit from 
dynamical variables, and is {\em not} prescribed from the outside. 
\item The next order of the Born-Oppenheimer scheme yields
 quantum gravitational correction terms proportional to $G$ \cite{KS,BK98,oup}.
The presence of such terms
leads to potentially observable effects in the
anisotropy spectrum of the cosmic microwave background (CMB)
radiation; see, for example, \cite{CK21} and the references therein. 
\end{itemize}
The formalism developed in \cite{BK98} is flexible enough to cope with
  a general background at the highest order of the Born--Oppenheimer
  approximation. That is, instead of the minisuperspace approximation
  used e.g. in \cite{CK21} at order $G^0$, one can use an arbitrary
  inhomogeneous solution of 
  the Hamilton--Jacobi equation. Ref. \cite{BK98} presents a full field-theoretic
  expansion.
  
The Born--Oppenheimer expansion scheme distinguishes a state of the form
(\ref{expiS}) from its complex conjugate. In fact, in a generic situation
where the total state is real, where it describes, for example, a superposition
of (\ref{expiS}) with its complex conjugate,
both states will decohere from each other, that is, 
they will become dynamically independent
\cite{deco}. This is a type of symmetry breaking in analogy to the
occurrence of parity violating states in chiral molecules. It is through
this mechanism that the i in the Schr\"odinger equation emerges.
Quite generally one can show how a classical geometry emerges
from quantum gravity in the sense of decoherence \cite{deco}: 
irrelevant degrees of freedom (such as density perturbations or
small gravitational waves) interact with the relevant ones
(such as the scale factor or the relevant part of the
density perturbations), which leads to quantum entanglement.
Integrating out the irrelevant variables (which are contained in the
above multipoles $\{ x_n\}$) produces a density matrix
for the relevant variables, in which non-diagonal (interference) terms
become small. One can show that the universe assumes classical
properties at the onset of inflation \cite{deco,oup}.

The recovery of the Schr\"odinger equation (\ref{semi}) raises an
interesting issue. It is well known that the notion of Hilbert space
is connected with the conservation of probability (unitarity) and thus
with the presence of an external time (with respect to which the
probability is conserved). The question then arises whether the concept
of a Hilbert space is still required in the {\em full} theory where
no external time is present. It could be that this concept makes sense
only at the semiclassical level where (\ref{semi}) holds, cf. our
remarks at the end of Section~3.2. 

Of course, the last word on quantum cosmology has not been spoken as
long as we have no consensus on the interpretation of the wave
function. What makes this issue so troublesome, is the missing link of
a wave function of the universe to measurement. The measurement
process supplies quantum mechanics with a probability
interpretation. The potentiality of measurement yields sense to the
Hilbert space structure. Expectation values are interpreted as
possible outcomes of measurements with probability depending on the
state the measured system is in. This interpretation entails the
normalizability requirement for the wave function. Moreover,
probabilities have to be conserved in time.

The problem is that we have no measurement crutch in quantum
cosmology. This is a problem that persists also in the full theory and
is a consequence of background independence. Only in a background of
space and time can we make observations. An expectation value
formulated in a theory deprived of that background is deprived of its
interpretation (and justification) through measurement.

A background independent quantum theory may thus be freed 
from a physical Hilbert
space structure.\footnote{This does not pertain 
the necessity to define an auxiliary or kinematical Hilbert space 
in order to define (not necessarily self-adjoint) operators.} 
It should keep linearity, since the superposition
principle is not linked to observation, but it should dismiss the inner
product as it is not clear how to endow it with a meaning
in a timeless context (a clear
statement on these issues is given in \cite{Vilenkin}).

A Hilbert-space structure may, however, be needed 
at an effective level for quantum
gravitational systems embedded in a semiclassical universe; a typical
situation is a quantum black hole \cite{paulo}: there, one has a
Schr\"odinger equation in which the time parameter $t$ is the WKB time
coming from the semiclassical universe, but where the Hamiltonian is
the indefinite Hamiltonian from the Wheeler--DeWitt equation as
applied to the black hole.

Due to the linear structure of quantum gravity, 
the total quantum state is a superposition of many macroscopic
branches even in the semiclassical situation,
each branch containing a corresponding version of the observer (the
various versions of the observer usually
do not know of each other due to decoherence). This is often referred
to as the `many-worlds (or Everett) interpretation of quantum theory`, although
only one {\em quantum} world (described by the full $\Psi$) exists.

\section{Arrow of time and structure formation}
\lb{Time}

Although most fundamental laws are invariant under time reversal,
there are several classes of phenomena in Nature that exhibit an
arrow of time \cite{Zeh}. It is generally expected that there is
an underlying master arrow of time behind these phenomena, and that this
master arrow can be found in cosmology. If there existed
 a special initial condition
of low entropy, statistical arguments could be invoked to demonstrate
that the entropy of the universe will increase with increasing size.

There are several subtle issues connected with this problem.
First, one does not yet know a general expression for the entropy
of the gravitational field; the only exception is the black-hole entropy,
which is given by the expression
\begin{equation}
\label{SBH}
S_{\rm BH}=\frac{k_{\rm B}c^3A}{4G\hbar}=k_{\rm B}\frac{A}{(2l_{\rm P})^2}
 \ ,
\end{equation}
where $A$ is the surface area of the event horizon, $l_{\rm P}$ is
the Planck length, and $k_{\rm B}$ denotes Boltzmann's constant.
According to this formula, the most likely state for our universe
would result if all matter would assemble into a gigantic black hole;
this would maximize (\ref{SBH}). More generally,
Roger Penrose has suggested to use the Weyl tensor as
a measure of gravitational entropy; see, for example,
\cite{Zeh,kiefer21} and the references therein. The 
cosmological situation is depicted in
Figure~5 which expresses the very special nature of the
big bang (small Weyl tensor) and the generic nature of a big crunch
(large Weyl tensor). Entropy would thus increase from big bang
to big crunch. 

\begin{figure}[h]
\label{fig_berlin05_7}
\caption{The classical situation for a recollapsing universe: the big crunch
is fundamentally different from the big bang because the big bang is
very smooth (low entropy), whereras the big crunch is very
inhomogeneous (high entropy).}
\begin{center}
\includegraphics[width=10cm]{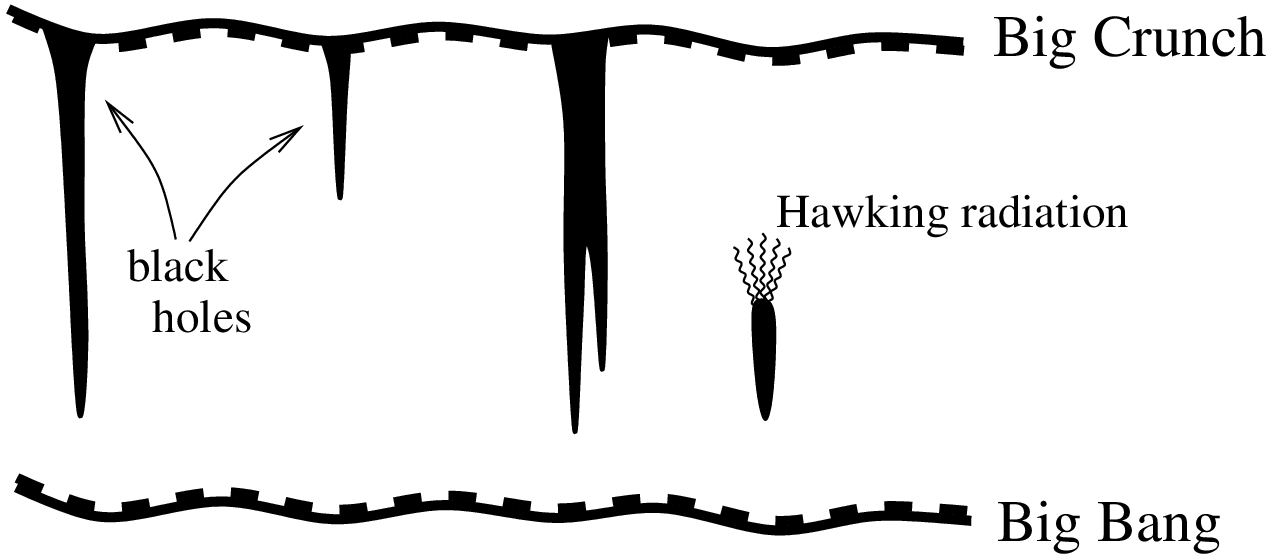}
\end{center}
\end{figure}

Second, since these boundary conditions apply in the very early 
(or very late) universe,
the problem has to be treated within quantum gravity.
But as we have seen, there is no external time in quantum gravity -- so
what does the notion `arrow of time' mean?

We shall address this issue in quantum geometrodynamics, but the
situation should not be very different in loop quantum cosmology or
string cosmology.
An important observation is that the
Wheeler--DeWitt equation exhibits a
fundamental asymmetry with respect to the `intrinsic time' defined
by the sign of the kinetic term. Very schematically, one can write this
equation as
\begin{equation}
 H \, \Psi = 
\left(\frac{\partial^2}{\partial\alpha^2} + \sum_i \, \left[
-\frac{\partial^2}{\partial x_i^2}+\underbrace{V_i(\alpha,x_i)}_{\to 0\ 
{\rm for}\ \alpha
\rightarrow -\infty}\right]\right) \, \Psi = 0 \ ,
\end{equation}
where again $\alpha=\ln a/a_0$ (and setting the reference scale
$a_0=1$), and the $\{ x_i\}$ again denote inhomogeneous 
degrees of freedom describing perturbations of the Friedmann universe
(see above); $V_i(\alpha,x_i)$ are the potentials of the inhomogeneities.
The important property of the equation is that the potential becomes small
for $\alpha\to -\infty$ (where the classical singularities would occur),
but complicated for increasing $\alpha$; 
the Wheeler--DeWitt equation thus possesses an asymmetry
with respect to `intrinsic time' $\alpha$. One can in particular impose
the simple boundary condition
\begin{equation}
\Psi \quad \stackrel{\alpha \, \to \, -\infty}{\longrightarrow}\ \psi_0(\alpha)\
\prod_i \psi_i(x_i)\ ,
\end{equation}
which would mean that the degrees of freedom are initially {\em not}
entangled. Defining an entropy as the entanglement entropy between
relevant degrees of freedom (such as $\alpha$) and
irrelevant degrees of freedom (such as most of the $\{ x_i\}$), this
entropy vanishes initially but
increases with increasing $\alpha$ because entanglement increases
due to the presence of the potential. In the semiclassical limit where
$t$ is constructed from $\alpha$ (and other degrees of freedom),
cf. (\ref{semi}), entropy increases with increasing $t$. This then
\emph{defines} the direction of time and would be the origin of
the observed irreversibility in the world. The expansion of the 
universe would then be a tautology. Due to the increasing entanglement,
the universe rapidly assumes classical properties for the
relevant degrees of freedom due to decoherence \cite{deco,oup}.
Decoherence is here calculated by integrating out the $\{ x_i\}$
in order to arrive at a reduced density matrix for $\alpha$.

\begin{figure}[h]
\label{fig_berlin05_8}
\caption{The quantum situation for a `recollapsing universe': big crunch
and big bang correspond to the same region in configuration space.}
\begin{center}
\includegraphics[width=10cm]{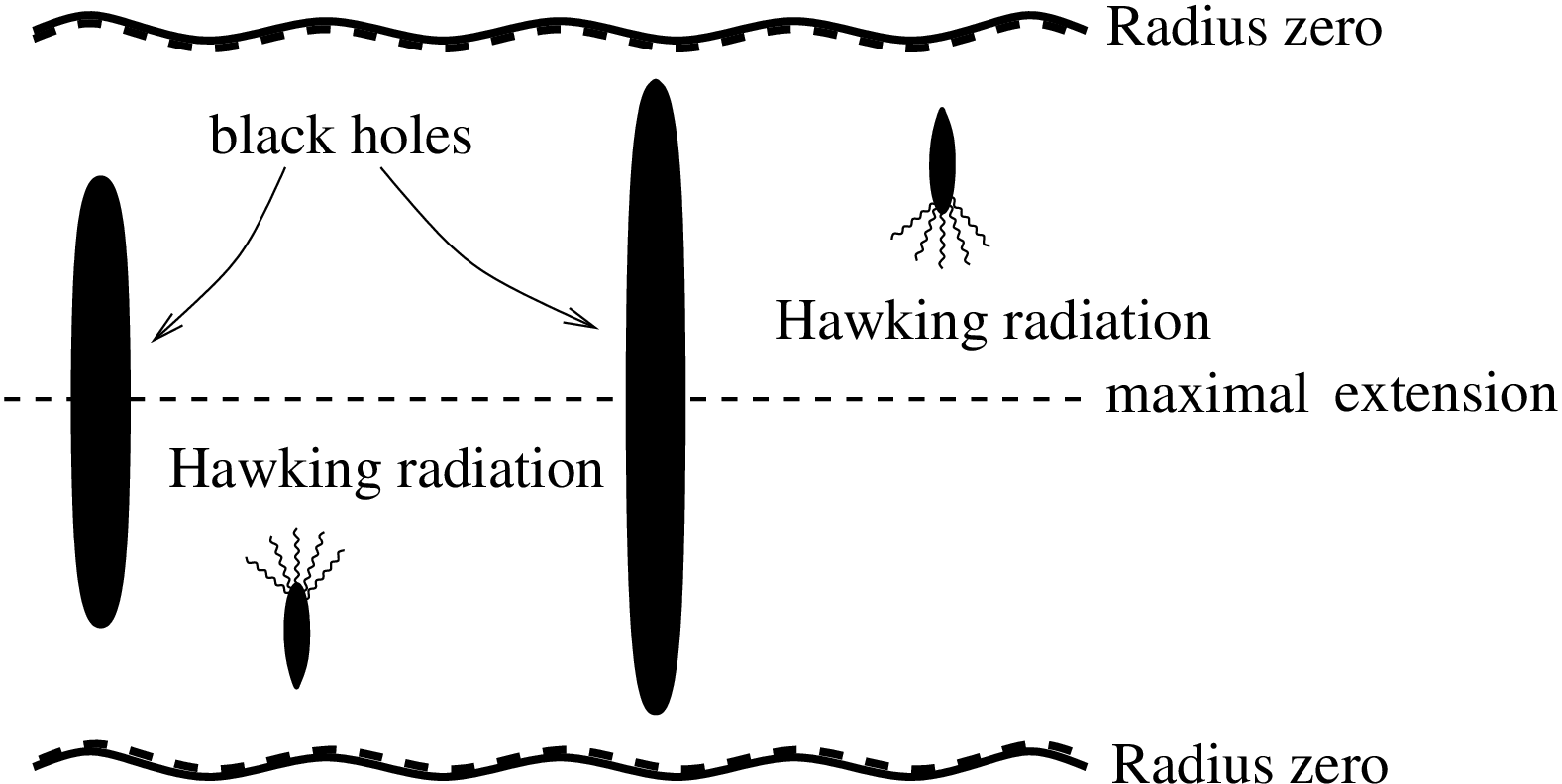}
\end{center}
\end{figure}

This process has interesting consequences for a classically
recollapsing universe \cite{KZ,Zeh}. 
Since big bang and big crunch correspond to the same region in
configuration space ($\alpha\to-\infty$), an initial condition
for $\alpha\to-\infty$ would encompass both regions,
cf. Figure~3. This would mean
that the above initial condition would always correlate increasing
size of the universe with increasing entropy: the arrow of time would
formally reverse at the classical turning point. 
Big bang and big crunch would be identical regions in
configuration space.
The resulting time symmetric picture is depicted in Figure~6,
which has to be contrasted with Figure~5.
As it turns out, however,
a reversal cannot be observed because the universe would enter a quantum phase
\cite{KZ}. Further consequences concern black holes in such a
universe because no horizon and no singularity would ever form.

These considerations are certainly speculative. 
They demonstrate, however, that interesting consequences would
result in quantum cosmology if the underlying equations were taken
seriously.

Once the background (described by the scale factor and some
other relevant variables) has assumed classical properties,
the stage is set for the quantum-to-classical transition of
the primordial fluctuations which serve as the seeds for structure formation.
The interaction with further irrelevant degrees of freedom
produces a classical behaviour for the field amplitudes of these fluctuations
\cite{KPS}.
These, then, manifest themselves in the form of classical stochastic 
fluctuations that leave their imprint in the anisotropy spectrum of
the cosmic microwave background radiation.


\section{A panacea for cosmological singularities? --- Singularity
  avoidance}
\lb{Singularity}
As we have discussed above,
a major motivation for the quantization of general relativity is the
occurrence of singularities in generic physical models (but see
  \cite{Mashoon,Konkowski} for alternative views).
This is why a touchstone of any quantum theory of gravity is its ability
to remove these singularities in some (still to be specified)
sense. In the cosmological context, the singularity of major concern
is the big-bang singularity. But restriction to
this singularity is far from exhausting even the possibilities
offered by the most plain homogeneous, isotropic models. Related to
dark energy is, for example, the big rip, a singularity occuring at
large scale factor \cite{MCB}.
 Another example of such a large-scale singularity
is the big brake \cite{KKS}, see the subsection below.

First one should make up ones mind about whether quantum
cosmology should resolve all conceivable singularities or just those
which are physically motivated in the most conservative sense (e.g. big
bang). Here it may help to recall the situation in quantum mechanics. The
classical divergence of the Coulomb potential is removed through
quantization. But nobody takes exception to the singularity persisting
in quantum mechanics for the potential which falls off with the
inverse squared distance, to mention one
example. In quantum
mechanics, singularity resolution is obviously restricted to the physically
relevant cases. Even though these cases are much harder to
make out in the cosmological context (as we are ignorant of the future
of our Universe and partly even of the past), one should keep this in
mind when discussing more exotic types of singularities.

Moreover, it is important to note that despite the very obvious
importance of a criterion for singularity avoidance, 
no formal criterion for the avoidance of
singularities at the quantum level exists by now. There are various
different notions but, so far, no systematic scheme has developed.
This lack of rigour is partly owed to an insufficient understanding
of singularities already at the classical level.
\subsection{Singularity avoidance of classical singularities defined through
  local criteria}
The criterion which is most important for the identification of a
cosmological singularity is the divergence of
curvature scalars. Metrics which produce infinite curvature scalars
are unphysical and therefore classified as singular. This is exactly
the criterion applying to the big-bang singularity. It is a
particularly useful criterion in quantum cosmology, for it relates the
singularity definition directly to properties of the metric. This allows
to mark singular regions in the space on which the wave function of
the universe is defined. 
In homogeneous, isotropic cosmologies this is the point where
the scale factor vanishes, $a=0$.
 From this, one arrives quite intuitively at the following criterion for
singularity avoidance at the quantum level:\\

\paragraph{Vanishing of the wave function}
The singularity is avoided if the wave function vanishes on
classically singular (in the sense that it produces diverging
curvature invariants) three-metrics.
This is perhaps the most intuitive and immediate way to think of
singularity resolution. It  may be the reason why it forms the
  content of the first boundary proposal in canonical quantum gravity
  put forward by DeWitt in 1967 \cite{DeWitt}. 
Actually, this idea comes to mind so
immediately that it is hard to support by arguments. Adding them
nonetheless, one would say that regions of configuration space on
which the wave function has no support are of no importance in quantum
theory as every statement deduced from quantum theory employs the wave
function.
Of course, this is again a generalization of what we know from ordinary quantum
theory. But it seems hard to argue that the role of the wave
function in quantum theory is completely altered by the inclusion of gravity to
the quantized interactions.\footnote{Note that this statement does not
touch upon the interpretation of the wave function nor upon the
question of how information may be retrieved from the wave
function. It just amounts to saying that the wave function is the
fundamental building block of any quantum theory. 
For an alternative point of view, see \cite{MB}.}
There are, in fact, 
several examples in the literature for singularity avoidance
through vanishing wave function; see, for example, \cite{Conradi,KKS}
as well as \cite{BKM19} and the references therein. 

In geometrodynamics, where the scale factor is restricted to be
positive, the big-bang singularity lies at the boundary of
configuration space. Vanishing of the wave function at the big bang is
therefore often due to a certain type of boundary condition, see Section
\ref{BoundaryConditions}. Even more, the choice of boundary conditions
might be justified from the strive for a singularity-free quantum
theory \cite{DeWitt,HH,Vilenkin}. (The situation is
  different in loop quantum cosmology where the big bang lies in the
  centre of configuration space due to the fact that here also
  negative values of the scale factor are allowed.)

But it must be emphasized that a non-vanishing, even a
diverging, wave function does not necessarily entail a singularity:
the ground-state wave function of the hydrogen atom (as found from the
Dirac equation), for example, diverges for $r\to 0$, but its norm
stays finite due to a factor $r^2$ in the measure. Vanishing of the
wave function can thus only be a sufficient, not a necessary criterion for
singularity avoidance. The DeWitt criterion can also be generalized to
take into account the conformal structure of superspace \cite{KKP19};
this is relevant for minisuperspaces of dimensions higher than two,
because for those cases the original DeWitt criterion is not
conformally invariant.

The DeWitt criterion is thus an intrinsic
principle of quantum cosmology. 
 It will be an important task to implement it 
in full quantum gravity with the same mathematical rigour
as the Hawking--Penrose singularity theorems are implemented in GR.

\paragraph{Well-defined quantum equations - Deterministic quantum evolution}
An obvious prerequisite for this condition to be satisfied is that the
equation describing the evolution of the wave function be well-defined
--- also at the singular metric. In the geometrodynamical picture this
is obtained through the introduction of the variable $\alpha=\ln a$ which
moves the big-bang singularity out to infinity. Loop quantum
  cosmology arrives at a finite evolution equation through a
  renormalization of the unbounded curvature terms. It is the fact
  that in loop quantum cosmology the big bang lies in the centre of
  configuration space that suggested to generalize this criterion to
  so-called `quantum hyperbolicity' \cite{MB,bojowald-book}. Quantum hyperbolicity
  asks for a deterministic evolution of general solutions to the constraint
  equations (Wheeler--DeWitt or the corresponding difference equation)
  through regions of classically singular metrics. 
\subsection{Singularity avoidance of classical singularities defined through
  global criteria}

Not all singularities arise through a divergence of curvature invariants.
In general relativity, 
a singularity is mathematically rigorously defined via geodesic
incompleteness. We can therefore speak of singularities (or their
avoidance) {\it only} when we have a notion of geodesics. This notion
is tightly knit to the concept of spacetime.
At the quantum-gravity level, spacetime is absent. We have three-metrics
and their canonically conjugate momenta but no prescription how to
stack these three-metrics together to obtain a four-dimensional
spacetime (due to the indeterminacy relations between the three-metric and its
conjugate momentum, which is related to the extrinsic curvature). 
The recovery of spacetime structure is possible only in the
semiclassical regime. Only here can we feasibly speak of spacetime
and only here can we apply the geodesic equation, thus deciding about
geodesic completeness or incompleteness.
Thus one arrives at the appealing picture that 
incomplete geodesics lead into quantum regimes --- 
their incompleteness being due to a break-down of the spacetime picture.\\
\paragraph{Break-down of the semiclassical approximation}
Singularity avoidance is here equivalent to the statement that the
semiclassical approximation breaks down in the region of the classical
singularities. This criterion can be found
in the early works by Hartle, Hawking and followers, see also
\cite{halliwell} and the discussion above.
 Here, regimes where the wave function is a real
exponential were
denoted as classically forbidden. Only where the wave function 
in the semiclassical approximation was
oscillatory, would one speak of a classically 
allowed region.
This argumentation follows closely our knowledge from
quantum mechanics where exponentially decaying wave functions occur in
classically forbidden regions.
Moreover, it is only out of oscillatory wave functions that one can
form wave packets. But only tightly peaked wave packets allow the
application of Ehrenfest's principle and the transition to a classical
picture. Consequently, a spreading of wave packets can also be
interpreted as a break-down of the semiclassical approximation 
\cite{oup,Zeh,CK88,MCB}.
\subsection{Example of singularity avoidance}

An illustrative model is the quantum version of a classical
cosmological model with a big-brake
singularity \cite{KKS}. A big-brake singularity is a singularity where
both scale factor and its first time derivative
stay finite, but the second derivative (the deceleration) diverges --
the universe comes to a halt infinitely fast. What makes this model
particularly interesting is the fact that the classical singularity 
occurs for big universes, that is, far away from the Planck scale
where the usual big-bang singularity occurs. 

The model describes a Friedmann universe with a scalar field that has
no mass but a potential of a special form. The classical trajectory in
configuration space is depicted in Figure~7.
\begin{figure}
\scalebox{1}{\hspace{-2mm}\includegraphics[angle=0]{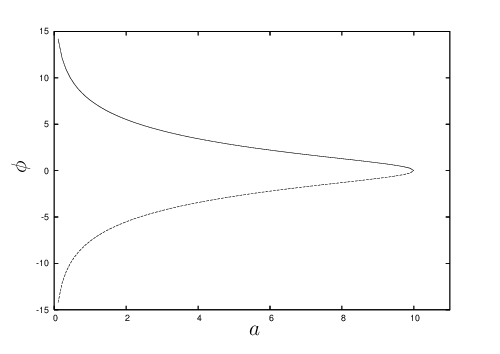}}
\caption{\lb{phi} Classical trajectory in configuration space
  \cite{KKS}. Degrees of freedom are the scale factor $a$ and scalar
  field $\phi$.
The singularity (big brake) is at $\phi=0$.}
\end{figure}
Upon discussing the Wheeler--DeWitt equation, one finds that all normalizable
solutions lead to a wave function that {\em vanishes} at the point of the
classical singularity; this we interpret as singularity avoidance. 
In Figure~8 we show a wave-packet solution that follows the classical
trajectory and that vanishes at the classical singularity.
\begin{figure}
\scalebox{0.6}{\hspace{-7mm}\includegraphics[angle=0]{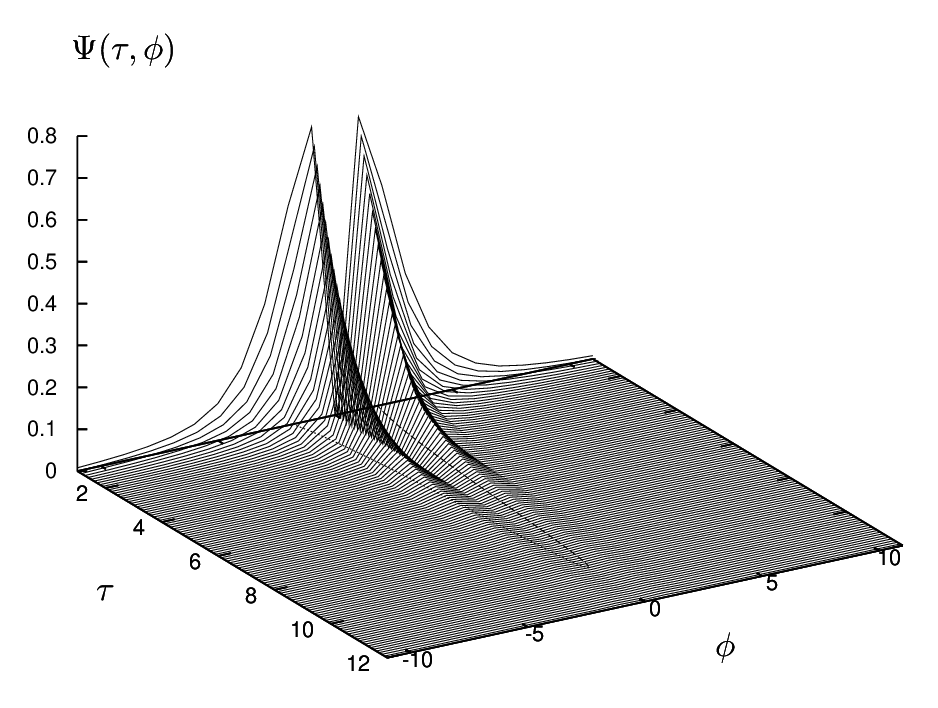}}
\caption{This plot shows the wave packet. 
It follows classical trajectories with initial values $a_0=1$ and
  $\phi_0\approx 0.88$. The classical trajectories are
depicted in the $(\tau,\phi)$-plane, with $\tau=a^6$. From \cite{KKS}.}
\end{figure}
An analogous situation of singularity avoidance is found in the loop
quantum cosmology of this model \cite{KKS}.
\section{Concluding remarks}
\lb{conclusion}
We have discussed that from a fundamental point of view it is important to deal with
 cosmology in quantum terms. This will cast light on many important
 issues such as the role of time, the origin and fate of the universe,
 the origin of irreversibility and structure, and the quantum
 measurement problem. 
All these were and are still pressing issues and
 opportunities which spur research in the field of quantum cosmology. 
Despite
 ongoing research and progress, many open questions remain. As the
 most important progress in quantum cosmology may consist in a better
 understanding of the questions and their relevance, we want to
 conclude our contribution by highlighting some of the most important
 questions in the field:

\begin{itemize}
\item How does one properly impose boundary conditions 
in quantum cosmology?
\item Is the classical singularity being avoided?
\item Will there be a genuine quantum phase in the future?
\item How does the appearance of our classical universe follow from
quantum cosmology?
\item Can the arrow of time be understood from quantum cosmology?
\item How does the origin of structure proceed?
\item Is there a high probability for an inflationary phase?
Can inflation itself be understood from quantum cosmology?
\item Can quantum-cosmological results be justified from full quantum
gravity?
\item Which consequences can be drawn from quantum cosmology 
for the measurement problem in quantum theory and for the field of
quantum information?
\item Can quantum cosmology be empirically tested?
\end{itemize}

We leave the answers for future research.

\section*{Acknowledgements}
We thank Alexander Vilenkin and Hongbao Zhang for their comments on
our manuscript.
B.S. thanks the Friedrich-Ebert-Stiftung for financial support. 
\begin{appendix}
\section{Derivation of the Wheeler--DeWitt equation for a concrete model}
\lb{appendix}
The starting point is the Einstein--Hilbert action of general relativity,
\begin{equation}
S_{\mathrm{EH}}=\frac{1}{2\kappa^2}\int \mathrm{d}^4x\sqrt{-g}(R-2\Lambda)\ ,
\end{equation}
where $\kappa^2=8\pi G$, $R$ is the Ricci scalar, $\Lambda$ denotes
the cosmological constant, $g$ stands for the determinant of the
metric and integration is carried out over all spacetime. The general
procedure of the canonical 
formulation and quantization was outlined in Section~2 and can be
found in detail in, for example, \cite{oup,GK}.

We want to apply the formalism for the case of a Friedmann universe,
which is of central relevance for quantum cosmology.  
In the homogeneous and isotropic setting the line element is given by
\be
\d s^2=-N^2(t)\d t^2+a^2(t)\d\Omega_3^2\ ,
\ee
where $\d\Omega_3^2$ is the line-element of an constant curvature space 
with curvature index $k=0,\pm 1$.

Here, $N$ is the lapse function, measuring the change of coordinate
time with respect to proper time. Setting $N=1$ yields the conventional
Friedmann time. This is a preferred foliation given for the isotropic
and homogeneous setting by observers comoving with the cosmological
perfect fluid, cf. Fig.~2.

The cosmological fluid shall here be mimicked by a scalar field $\phi$ 
 with potential $V(\phi)$.
The model is thus described by the action
\ben
S&=&S_{\mathrm{grav}}+S_{\mathrm{matter}}\nonumber\\
&=&\frac{3V_0}{\kappa^2}\int \d
tN\left(-\frac{a\dot{a}^2}{N^2}+ka-\frac{\Lambda
    a^3}{3}\right)\nonumber\\ & &  
+\frac{V_0}{2}\int \d t
Na^3\left(\frac{\dot\phi^2}{N^2}-2V(\phi)\right)\ .
\een
Here, $V_0$ denotes the (dimensionless) coordinate volume of a spatial slice.
In the following we choose, for convenience, $V_0=2\pi^2$ (assuming
the three-space is a three-sphere).
We thus have a Lagrangian system with two degrees of freedom, $a$ and
$\phi$. 
The canonical momenta read
\be
\lb{momenta}
\pi_a=\frac{\partial L}{\partial \dot{a}}=-\frac{6a\dot{a}}{\kappa^2N}\ , \quad \pi_{\phi}=\frac{\partial L}{\partial \dot{\phi}}=\frac{
a^3\dot{\phi}}{N}, \quad \pi_N=\frac{\partial L}{\partial \dot{N}}\approx 0
\ .
\ee
Curly equal signs stand for weak equalities, that is, equalities which
hold after the equations of motion are satisfied. The equation for
$\pi_N$ is thus a primary constraint.
The Hamiltonian reads
\ben
\lb{constraint}
{\mathcal
  H}&=&\pi_a\dot{a}+\pi_{\phi}\dot{\phi}+\pi_N\dot{N}-{\mathcal L}\nonumber\\
&=&-\frac{\kappa^2}{12a}\pi_a^2+\frac{1}{2}\frac{\pi_{\phi}^2}
{a^3}+a^3\frac{\Lambda}{\kappa^2} + a^3V-\frac{3ka}{\kappa^2}\ .
\een 
As we assumed homogeneity and isotropy, the diffeomorphism constraints
are satisfied trivially.
 From the preservation of the primary constraint $\pi_N\approx 0$ one
finds that the Hamiltonian is
constrained to vanish, ${\mathcal H}\approx 0$. 
Expressed in terms of the `velocities', $\dot a$ and $\dot\phi$,
this constraint becomes identical to the Friedmann equation,
\be
\lb{Friedmann}
\left(\frac{\dot{a}}{a}\right)^2\equiv H^2=
\frac{\kappa^2}{3}\left(\frac{\dot{\phi}^2}{2}+V(\phi)\right)
+ \frac{\Lambda}{3}-\frac{k}{a^2}\ .
\ee

The space spanned by the canonical variable (the three-metric)
 in the full theory is called superspace, see the main text.
 In dependence on this denotation, the space spanned
by $(a,\phi)$ is called minisuperspace. The metric on this
space is named after DeWitt and is for this model given by
\be
G_{AB}=\left(\begin{array}{rl}-\frac{6a}{\kappa^2}
&\quad 0\\0&\quad a^3\end{array}\right)\ .
\ee
Note the indefinite nature of the metric, resulting in an indefinite
DeWitt metric for general relativity. The
Hamiltonian constraint in this minisuperspace model thus reads
\be
{\mathcal H}=N\left(\frac{1}{2}G^{AB}\pi_A\pi_B+{\mathcal V}
(q)\right)\approx 0\ ,
\ee
where 
\be
{\mathcal V}(q)=\frac12\left(-\frac{6ka}{\kappa^2}
+\frac{2\Lambda
    a^3}{\kappa^2}+a^3V(\phi)\right)\ ,
\ee
$A,B=\{a,\phi\}$, that is, $q^1=a, q^2=\phi$, and $G^{AB}$ is the
inverse DeWitt metric.
 It is an artefact of the two-dimensionality
of the model considered here that minisuperspace is conformally flat. 
In more complicated settings, minisuperspace can also be curved and
additional curvature terms may occur depending on the choice of 
factor ordering \cite{KKP19}.

Dirac's constraint quantization requires an implementation of
(\ref{constraint}) as 
\be
\hat{{\mathcal H}}\Psi=0\ .
\ee 
The operator $\hat{{\mathcal H}}$ is constructed from the
conventional Schr\"odinger representation of canonical variables,
\be
\hat{q}_A\Psi\equiv q_A\Psi\ , \qquad\qquad
\hat{\pi}_A\Psi\equiv\frac{\hbar}{\mathrm i}
\frac{\partial}{\partial q_A}\Psi\ ,
\ee 
for $A\in\{a,\phi\}$. The Hamiltonian operator is, of course, not
uniquely defined in this way. On the contrary, factor-ordering
ambiguities occur here as in ordinary quantum mechanics --- with the
important difference that here factor ordering cannot be justified by
experiment. Usually one decides on the covariant ordering, the
Laplace--Beltrami ordering,
\be
G^{AB}\pi_A\pi_B\rightarrow
-\hbar^2\bigtriangledown_{\mathrm{LB}}^2=-\frac{\hbar^2}{\sqrt{-{\mathcal
      G}}}
\partial _A\left(\sqrt{-{\mathcal G}}G^{AB}\partial _B\right)\ ,
\ee
where ${\mathcal G}$ denotes the determinant of the DeWitt--metric.
Choosing
Laplace--Beltrami factor ordering, the Wheeler--DeWitt equation reads
\be
\lb{wdw1}
\left(\frac{\hbar^2\kappa^2}{12}a\frac{\partial}{\partial a}
a\frac{\partial}{\partial a}-\frac{\hbar^2}{2}\frac{\partial^2}
{\partial\phi^2}+a^6\left(V(\phi)
+\frac{\Lambda}{\kappa^2}\right)-\frac{3ka^4}{\kappa^2}\right)
\Psi(a,\phi)=0\ .
\ee
Introducing $\alpha\equiv\ln a/a_0$, with $a_0$ as reference scale,
one obtains the following equation, which is (\ref{whdw1}) of the main
text, 
\be
\lb{wdw2}
\left(\frac{\hbar^2\kappa^2}{12}\frac{\partial^2}{\partial\alpha^2}-
\frac{\hbar^2}{2}\frac{\partial^2}{\partial\phi^2}+a_0^6
e^{6\alpha}\left(V\left(\phi\right)+\frac{\Lambda}{\kappa^2}\right)-3a_0^4e^{4\alpha}\frac{k}{\kappa^2}\right)\Psi(\alpha,\phi)=0 
\ .
\ee
This is an equation of the same form as the Klein--Gordon equation: 
the derivative with respect to $\alpha$ corresponds to a time
derivative, the derivative 
with respect to $\phi$ to a spatial derivative, and the remaining terms constituting a `time and space dependent' mass term, 
that is, a non-trivial potential term. Equation (\ref{wdw2}) is the
starting point for many discussions in quantum cosmology. The physical
units are often chosen to be $\kappa^2=6$ for convenience.
\end{appendix}
%
%


\end{document}